\documentclass[aps,pre,twocolumn,groupedaddress,amssymb]{revtex4}
\usepackage{graphicx}
\usepackage[figuresright]{rotating}
\begin{document}

\def\mdr{\mathit{\delta r}}
\newcommand{\aver}[1]{\langle #1 \rangle}

\title{Test of multiscaling in DLA model using an
off-lattice killing-free algorithm}

\author{Anton Yu. Menshutin and Lev N. Shchur}
\affiliation{Landau Institute for Theoretical Physics, 142432 
Chernogolovka, Russia\\
and Moscow Institute of Physics and Technology (MFTI), 141700 
Dolgoprudny, Russia}
\begin{abstract} 

We test the multiscaling issue of DLA clusters using a modified
algorithm. This algorithm eliminates killing the particles at the
death circle. Instead, we return them to the birth circle at a
random relative angle taken from the evaluated distribution. In
addition, we use a two-level hierarchical memory model that allows using
large steps in conjunction with an off-lattice realization of the
model. Our algorithm still seems to stay in the framework of the
original DLA model. We present an accurate estimate of the fractal
dimensions based on the data for a hundred clusters with 50 million
particles each.  We find that multiscaling cannot be ruled out. We
also find that the fractal dimension is a weak self-averaging
quantity. In addition, the fractal dimension, if calculated using the
harmonic measure, is a nonmonotonic function of the cluster radius.  We
argue that the controversies in the data interpretation can be due to
the weak self-averaging and the influence of intrinsic noise.

\end{abstract}


\maketitle
\section{Introduction}

The DLA model~\cite{WS} plays the same role 
in the physics of structure growth in
two-dimensions~\cite{BH} as the Ising model plays in the theory 
of phase transitions. It catches the
main features of the random growth and is quite simple in
definition. But this model however is still not solved analytically
more than twenty years after
its introduction~\cite{WS}. Direct simulations
of the original model and calculations using the conformal-mapping
technique~\cite{HL} are the two main methods for investigating DLA
structures.

It is commonly believed that DLA clusters are random
fractals~\cite{WS,BH}, and the accepted estimate for the fractal
dimension is $D=1.71\pm 0.01$. The analytic result~\cite{Hastings}
predicts $D=17/10$ in agreement with the numerical estimates.  
The surface of a
DLA cluster demonstrates the multifractal properties obtained in
simulations~\cite{MSCW,HMP} and supported analytically~\cite{HDH}.

Several groups claim DLA clusters have multiscaling 
properties~\cite{ms-PZ,ms-DHOPSS,ms-CZ,ms-ACMZ,ms-Os91}:
that the penetration depth $\xi$ is scaled differently from the
deposition radius $R_{\rm{dep}}$ and that a whole set of scaling exponents
exists within the framework of multiscaling. Recently, these claims
were doubted in papers by Somfai, Ball, Bowler, and
Sander~\cite{nms-BBSS,nms-SBBS}.

The off-lattice killing-free algorithm, our implementation of
the DLA algorithm, allows generating a large number of huge clusters
and calculating the fractal dimensions of the quantities mentioned above.
Our numerical results do not support the arguments presented 
in~\cite{nms-BBSS,nms-SBBS} but favor  the early
results~\cite{ms-PZ,ms-DHOPSS,ms-CZ,ms-ACMZ,ms-Os91}.
 
The multiscaling that was ``suspected'' in those papers was attributed
in~\cite{nms-BBSS,nms-SBBS} to the strong lattice-size effects they
advocated, with the correction-to-scaling exponent $1/3$. We do not
find any evidence for that in our data. Instead, we prefer to
attribute the spreading of the fractal dimension values as estimated from
the different quantities to the weak-self-averaging of the fractal
dimension. Its relative fluctuation decays with the number of
particles approximately as ${\cal
F}_D=(\aver{D^2}-\aver{D}^2)/\aver{D}^2)\propto 1/N^{\gamma}$ with
$\gamma\approx 0.35\pm 0.04$ for the large cluster sizes, i.e, with
the exponent about three times smaller than would be expected for the
usual averaging of the quantity. Thus one can expect large
fluctuations of the fractal dimension as estimated from the different
quantities, with different methods, and from different cluster 
sizes.

The alternative and more naive view is to say that the last exponent
is also evidence for the multiscaling properties of DLA cluster.

Our algorithm is off-lattice with memory organization similar to one
used in the Ball and Brady algorithm~\cite{BB-alg}. The main differences are:
 (i) we use only two layers in the memory hierarchy, which seems
optimal for reducing memory usage, keeping the overall
efficiency of simulations high; (ii) we use bit-mapping for the second layer
of memory to reduce the total memory used by the simulation program;  
(iii) we use large walk steps. The new feature is the recursive
algorithm for the free zone tracking.

The essential feature of our algorithm is that we modified the rule
for the particles that go far from the cluster. We never
kill any particle at the outer circle but return them to the birth
circle~\cite{SSZ,KVMW} with an evaluated probability.  
This procedure eliminates the effect of the potential distortion (and thus
of the cluster harmonic measure), keeping our algorithm in the same
universality class.

In this paper  we present all details of our algorithm. We 
believe that some details may be important both for understanding
the results and for comparing the results properly.  
We briefly discuss the essentials of our implementation of the
off-lattice killing-free algorithm in Section~\ref{sec-alg},
accompanied with some technical details on the derivation of the
return-on-birth-circle probability given in Appendix~\ref{sec-prob}
and details on the memory model in Appendix~\ref{sec-mem}. 
Various methods for estimating the fractal
dimension are described in 
Section~\ref{sec-fractal} and compared with those from other
off-lattice estimations.  We discuss fluctuations 
of the fractal dimension 
in Section~\ref{sec-self} and the multiscaling issue in
Section~\ref{sec-multi}. The discussion in Section~\ref{sec-sum} 
concludes our paper.

\section{Killing-free Off-lattice Algorithm}
\label{sec-alg}

A cluster grows according to the following rules:
(1) We start with the seed particle at the origin.
(2) A new particle is born at a random point on the circle of
radius $R_{\rm b}$.
(3) A particle moves in a random direction; the step length is chosen as
big as possible to accelerate simulations.
(4) If a particle walks out of the circle of radius 
$R_{\rm d}>R_{\rm b}$, it is returned
to the birth radius $R_{\rm b}$ at the angle $\varphi'$ taken from 
distribution~(\ref{prob-norm}) and relative to the last
particle coordinate.
(5) If particle touches a cluster, it sticks.
(6) The cluster memory is updated. Steps from (2) through (6) 
are repeated $N$ times.

The difference from the traditional DLA algorithm is in Step 4. We
remind the reader that in the original algorithm, a particle is killed
when it crosses $R_{\rm d}$ and the new one starts a walk from a random
position on the circle $R_{\rm b}$, i.e, at the position
$(R_{\rm b},\varphi_{random})$. Clearly, the rule being perfect in the limit
$R_{\rm d}\rightarrow\infty$ will influence the growth stability when 
$R_{\rm d}$ is finite.

Instead, if particle crosses $R_{\rm d}$ to a position with $r>R_{\rm d}$,
we use
the probability~\cite{SSZ,KVMW} as determined by expression~(\ref{prob-norm}) in
Appendix A with $x=r/R_{\rm b}$ to obtain new particle position.
The particle then walks 
from the position $(R_{\rm b},\varphi'+\varphi)$, assuming $(r,\varphi)$
is the old particle position. Details on obtaining expression~(\ref{prob-norm})
and on generating random numbers with given probability 
are presented in~\cite{SSZ} and in Appendix A.

During the cluster growth, the empty space between cluster branches
also grows. A special organization of the memory is implemented 
to avoid long walks in the empty space. We modify the
hierarchical memory model~\cite{BB-alg}, using only a two-layer hierarchy
(see Appendix~\ref{sec-mem} for details) and a bit-mapping technique
for the second layer to reduce total memory.

\begin{figure} 
\centering
\includegraphics[angle=0,width=\columnwidth]{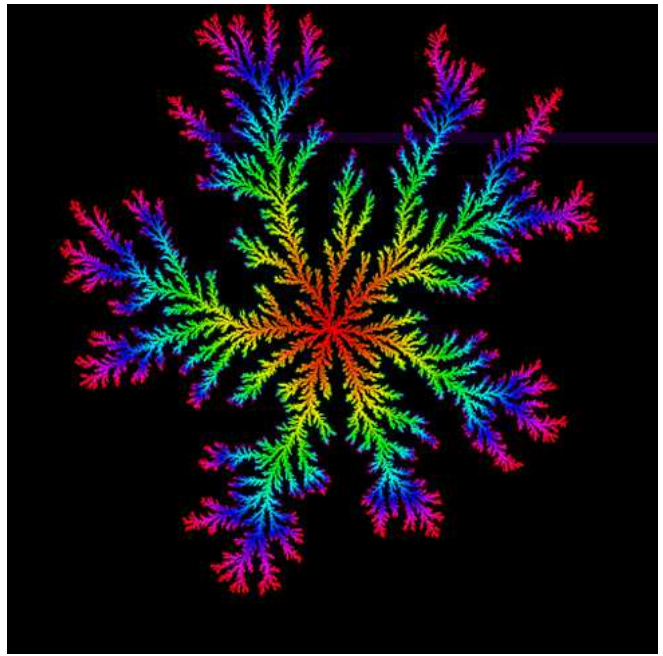}
\caption{(Color online) Typical DLA cluster with $5\cdot10^7$ particles
grown using killing-free off-lattice algorithm. Color denotes particle age.} 
\label{fig-typical}
\end{figure}

\section{Fractal dimension estimations}
\label{sec-fractal}

In this section we present the results on the estimation for the DLA
cluster fractal dimension analyzing various cluster lengths: 
deposition radius $R_{\rm dep}$, mean square radius $R_2$, gyration radius 
$R_{\rm gyr}$, effective radius $R_{\rm eff}$ and ensemble penetration depth
$\xi$. The dependence of the length $R$ on the number of particles $N$ gives
estimation of the fractal dimension through the relation $R\propto
N^{1/D}$. There are two ways (see Table~\ref{tab-fractal}) to extract
this dependence. The first is to average over the ensemble of clusters,
for example, $\aver{r(N)}=\sum_{i=1}^{K}r_i(N)/K$, where the sum is over
$K$ clusters and $r_i(N)$ is the position of the $N$-th particle in the $i$-th
aggregate. The second is to average over the harmonic measure, which is the
probability of sticking at the point $r$, 
for example, $R_{\rm dep}=\int r\,{d}q$.

\begin{table} \begin{tabular}{|l|l|l|l|l|}
 & Definition 1 & $D$ & Definition 2 & $D$  \\
\hline
$R_{\rm dep}$ & $\aver{r} $  & $1.70942(46)$  & $\aver{ \int r\, {d}q}$& 
$1.70922(97)$ \\ 
$R_2$ & $\sqrt{\aver{r^2}}$ & $1.71003(45)$ &$\aver{\sqrt{ \int r^2\, {
d}q}}$ & $1.7087(11)$ \\
$R_{\rm gyr}$ & $\sqrt{\frac1N \sum_{k=1}^N 
\aver{r^2}_k}$&$1.71008(96)$&---&--- \\
$R_{\rm eff}$ &--- & ---& $\aver{\exp{(\int \ln r\, {d}q)}}$& 
$1.70944(87)$\\
$\xi$ & $\sqrt{{R_2}^2-{R_{\rm dep}}^2}$&$1.74(3
)$ &---&$1.69(7)$
\end{tabular}
\caption{Estimates of the fractal dimension $D$ extracted with the fit
$N \propto R^{D}$ to the dependence of the various lengths $R$ (
deposition radius $R_{\rm dep}$, mean square radius $R_2$,  gyration radius 
$R_{\rm gyr}$, effective radius $R_{\rm eff}$ and  ensemble penetration
depth $\xi$)  on the number of particles $N$. The third (fifth)  
column is the fit to the data calculated with the definition given in the
second (fourth) column.}
\label{tab-fractal}
\end{table}

We use data from $K=100$ clusters, each built up by the algorithm
described in Section~\ref{sec-alg} with $5\cdot10^7$
particles.  A typical cluster~\footnote{a high quality figure is
available at http://www.comphys.ru/may/dla-cluster.eps} is shown in
Figure~\ref{fig-typical}. Estimations of the fractal dimension from the
various cluster lengths using both ways of averaging are shown in
Table~\ref{tab-fractal}. The errors given in parentheses as
corrections to the last digit include both statistical errors and
fitting errors.

In averaging over the harmonic measure (the last column in
Table~\ref{tab-fractal} with the preceding definition column), we first
estimate $D$ from a single cluster~\footnote{Averaging over the harmonic
measure in practice is done by freezing the growth and using probe
particles that walk according to the usual growth algorithm but do not stick
to the cluster. Then a new particle starts its motion. 
We use $10^4$ probe particles typically. The average of the
positions $r$ where they touch the cluster gives the quantity averaged
over the harmonic measure.} and than average over the $K=100$ samples.

The estimations from all lengths agree well with 
each other within the error and with the most
accepted value of the fractal dimension $D=1.711$ (see, e.g.~\cite{nms-BBSS}).
The error in measuring  
the penetration depth is much higher because of its complex structure.

It was proposed in~\cite{nms-BBSS,nms-SBBS} that the various lengths
depend on the number of particles with the correction term
\begin{equation}
R(N)=\hat{R}N^{1/D}(1+\tilde{R}N^{-\nu}),
\label{correction}
\end{equation}
\noindent and that exponent $\nu$ is the same 
($\nu=0.33$) for all quantities. 

The fit of the data to expression~(\ref{correction}) with the fixed values 
of $D=1.711$ and $\nu=0.33$ is presented in Table~\ref{tab-cor}. 
Table~\ref{tab-cor} here should be compared with Table~1 
in~\cite{nms-SBBS}. The difference in the values of $\hat{R}$ is about 
factor of two and probably because of the different units of the particle 
size used in simulations. We fix to unity the particle radius and not the 
particle diameter. Result of the fit is extremely sensitive to the value 
of $D$ used. For example, if we fix fractal dimension $D$ to the value 
$1.710$ which one may suggest from our Table~\ref{tab-fractal}, the values 
of $\hat{R}$ and $\tilde{R}$ for the fit to $R_{gyr}$ changed from those 
in the third line of our Table~\ref{tab-cor} to $0.958(1)$ and $-0.06(1)$. 
Thus, values of $\hat{R}$ differ by six standard deviations, and values of 
$\tilde{R}$ -- by sixteen standard deviations.

If we fit our data for the different cluster realizations to the 
expression~(\ref{correction}) without fixing $D$ and $\nu$, then we find a 
large fluctuation of $\tilde{R}$ around the zero value.

If we suppose that authors of Ref.~\cite{nms-SBBS} fixed diameter to 
unity, we may conclude that our data for $\hat{R}$ from 
Table~\ref{tab-cor} coincide with the corresponding data in 
Table~ of 
Ref.~\cite{nms-SBBS} and not the data for $\tilde{R}$. We can therefore 
conclude that the results of fitting to expression~(\ref{correction}) are 
inconclusive and that the values of the coefficient $\tilde{R}$ presented 
in Table~\ref{tab-cor} are just random.

These large fluctuations can be understood in the framework of the
weak self-averaging of the fractal dimension, which we describe in the next
section.

\begin{table} 
\begin{tabular}{|l|l|l||l|l|} 
& \multicolumn{2}{|c|}{Definition 1}   & \multicolumn{2}{|c|}{Definition 2} \\ \hline
 & $\hat{R}$& $\tilde{R}$  & $\hat{R}$ & $\tilde{R}$\\ \hline
$R_{\rm dep}$ & 1.394(2) & $0.59\pm 0.28$  & 1.398(2) & $0.006 \pm 0.030$ \\
$R_2$     & 1.414(2) & $0.22\pm 0.27$  & 0 & 0 \\
$R_{\rm gyr}$ & 0.964(1) & $-0.22\pm 0.01$ & 0 & 0 \\
$\xi$     & 0.239(1) & $-13\pm 1$      & 0 & 0 
\end{tabular}
\caption{Coefficients of correction to scaling fits (expr.~\ref{correction})
with fixed $D=1.711$ and $\nu=0.33$. The definitions of various
length $R$ are given in Table~\ref{tab-fractal}.}
\label{tab-cor}
\end{table}

\section{Weak self-averaging of $D$}
\label{sec-self}

We check how the fluctuations of the measured fractal dimension
$D$ depend on the system size $N$. By analogy with thermodynamics,
the relative fluctuation ${\cal 
F}_D=\aver{D^2}-\aver{D}^2)/\aver{D}^2$ of the quantity $D$ should
decrease as the inverse system size. For full
self-averaging of the fractal dimension,   ${\cal
F}_D\propto 1/N \propto 1/R^D$ can be expected.
 In the case of a slower decay ${\cal
F}_D\propto 1/N^{\gamma}$ with $\gamma<1$, one can say that
self-averaging of $D$ still occurs and that it is weak self-averaging.

\begin{figure} \centering
\includegraphics[angle=0,width=\columnwidth]{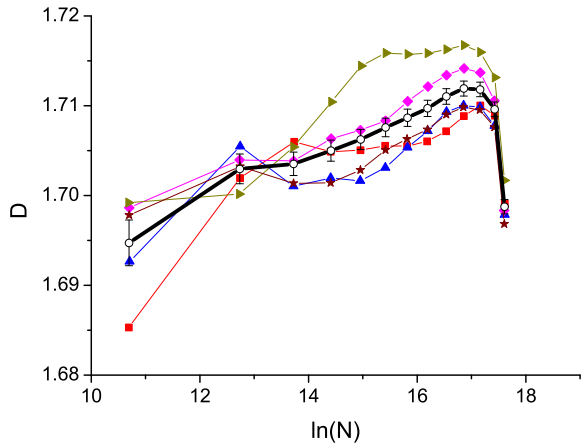} 
\caption{(Color online) Fractal dimension $D(N)$ as a function of the number $N$ of
particles inside the radius $R$. Solid curve with error bars 
represents $D(x)$ averaged over 100 clusters. Another curves are the
averages over five different 20-clusters ensembles.}
\label{fig-d-simple} 
\end{figure}

We extract the fractal dimension from the analysis of the clusters in two 
ways. First, we calculate the number $N$ of particles inside the circle of 
radius $R$ for the given cluster. The slope of this curve on a log-log 
plot gives $D_i$. The fractal dimension $D_i$ as a function of $N$ is 
denoted as $D_i(N)$, where $i$ is the number of cluster, $i=1,2,...,K$, 
and $K=100$. Then we average $D_i(N)$ over the ensemble of $K$ clusters, 
$D(N)=\frac1K\sum_{i=1}^K D_i(N)$. The fractal dimension $D(N)$ is plotted 
in Figure~\ref{fig-d-simple} with a bold line as a function of the system 
size $N$. To better understand the behavior of $D$, we also present the 
results of averaging over smaller ensembles. We divide the whole ensemble 
with 100 clusters into five independent groups. Averaging over each group 
gives five different curves for $D(N)$, which exhibit strong fluctuations. 
Error bars are computed as fluctuations of $D_i(N)$ in the ensemble of 
$K=100$ clusters.

For sufficiently large $N>10^5$, the values of $D(N)$ vary mainly in the range
1.695--1.715, which is about the usually accepted value of the fractal
dimension. The drop off of the curve $D(N)$ at $N>2\cdot 10^7$ is due
to the influence of the cluster boundary:  the most active zone of 
cluster growth, which is underdeveloped in comparison with the rest
of the cluster, is now inside $R$.

\begin{figure}
\centering
\includegraphics[angle=0,width=\columnwidth]{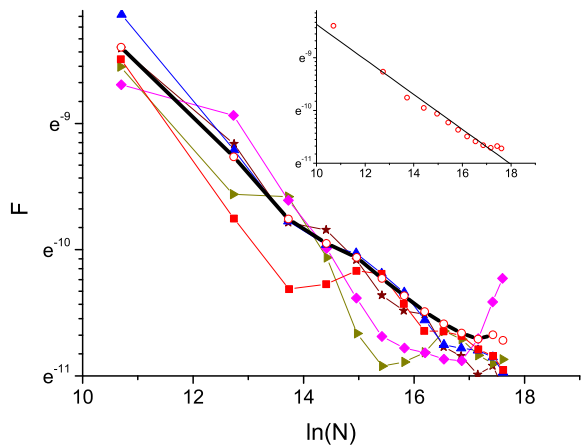}
\caption{(Color online) Decay of the relative fluctuations of the fractal 
dimension shown in Figure~\ref{fig-d-simple}. The symbols are the same as 
in Figure~\ref{fig-d-simple}. The solid line in the inset is the linear fit 
to the open circles.}
\label{fig-var-D-single}
\end{figure}

Relative fluctuations ${\cal F}_D$ of that quantity are shown in 
Figure~\ref{fig-var-D-single}. The bold line represents 100-cluster 
ensemble averaging compared with five lines computed using five 20-cluster 
groups. Fluctuations decrease with the exponent $\gamma=0.33\pm 0.02$.  
This is three times slower decreasing than expected in the case of full 
self-averaging.

\begin{figure}
\centering
\includegraphics[angle=0,width=\columnwidth]{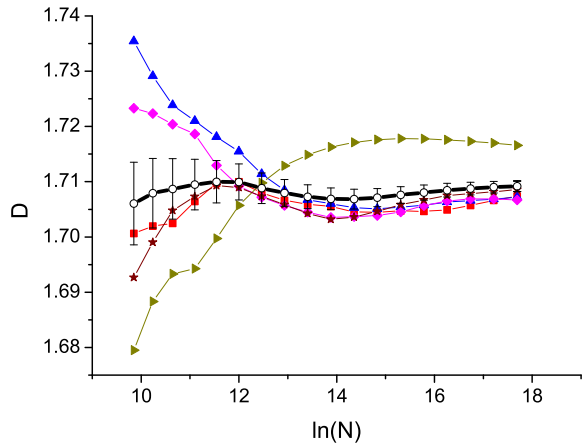}
\caption{(Color online) Fractal dimension $D_{\rm harm}(N)$ as a function of the
number of particles in the cluster used to estimate $R_{\rm dep}$
with averaging over the harmonic measure. The symbols are the same as
in Figure~\ref{fig-d-simple}.}
\label{fig-d-harm}
\end{figure}

Next, we analyze self-averaging of the fractal dimension as extracted
from the dependence of the deposition radius $R_{\rm dep}(N)$ calculated
by averaging over the harmonic measure. This quantity averaged over the
ensemble of $100$ clusters is plotted in Figure~\ref{fig-d-harm} as a
function of the system size $D_{\rm harm}(N)$. It exhibits some
``oscillation'' around the value 1.708, which is very close to the
accepted DLA fractal dimension. Averages over smaller ensembles also
demonstrate this feature. It is not clear how this quantity will
change  as the system size increases further.

\begin{figure}
\centering
\includegraphics[angle=0,width=\columnwidth]{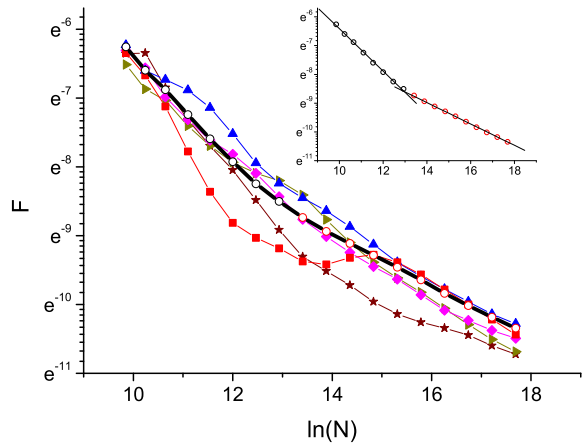}
\caption{(Color online) Decay of the relative fluctuations of the fractal 
dimension $D_{\rm harm}(N)$ shown in Figure~\ref{fig-d-harm}.
The symbols are the same as in Figure~\ref{fig-d-simple}. Inset: 
the linear fit to the data with $N$ up to $5\cdot 
10^5$ and the linear fit to the data with larger $N$.}
\label{fig-var-D-harm}
\end{figure}

Relative fluctuations ${\cal F}_{D_{\rm harm}}$ of $D_{\rm harm}$ are
shown in Figure~\ref{fig-var-D-harm}. There are two regimes of the
${\cal F}_{D_{\rm harm}}$ decay. First, for the cluster sizes $N<10^6$, it
decays with $\gamma=0.71\pm 0.02$, much faster than for the
${\cal F}_D$ as estimated with the conventional counting method
described above with $\gamma \approx 0.33$. The next regime, for the
larger system sizes $N>10^6$, shows slower decay with the exponent
$\gamma=0.38\pm 0.01$, close to those estimated by the traditional
counting method~\footnote{We note that the close value of the
exponent $\gamma$ was estimated for the quasi-flat growth of a DLA cluster
using Hastings--Levitov conformal mapping simulations~\cite{RS}.}.

In practice, the exponent value $\gamma \approx 0.33$ means that 
to obtain more accurate estimate of the fractal dimension,
one have to increase the number of particles in the cluster by
three orders of magnitude.

\section{Multiscaling}
\label{sec-multi}

\begin{figure} \centering
\includegraphics[width=\columnwidth]{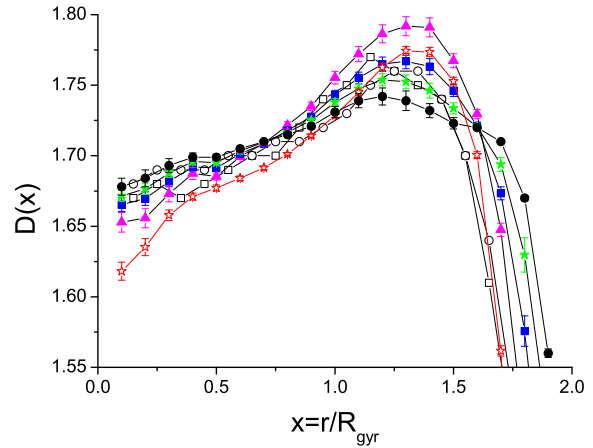}
\caption{(Color online) Multiscaling fractal dimension D(x) for different cluster 
sizes: solid triangles, $N=10^6$; solid squares, $N=10^7$; stars, 
$N=5\cdot 10^7$; solid circles, limit of our data for $N\rightarrow 
\infty$; open symbols are from~\cite{ms-ACMZ}; open  squares, $N=10^4$ 
square lattice; open circles, $N=10^5$ off-lattice.} 
\label{d_x}
\end{figure}

It was long discussed that DLA clusters are not simple
fractals~\cite{ms-PZ,ms-DHOPSS,ms-CZ,ms-ACMZ,ms-Os91} and, for
example, that the fractal dimension depends continuously on the normalized
distance from the cluster origin~\cite{ms-ACMZ}.  It was suggested
in~\cite{ms-ACMZ} that the density of particles at a distance $r$ from
the origin obeys the equation $g(r,R_{\rm gyr})=c(x)R_{\rm gyr}^{D(x)-d}$, 
where $x=r/R_{\rm gyr}$, with a nontrivial (non constant)
multiscaling exponent $D(x)$ (lines with open symbols in
Figure~\ref{d_x}).

Quite recently, Somfai et al.~\cite{nms-BBSS,nms-SBBS} claimed that
the multiscaling picture is wrong and is misled by finite-size transients.
They argued for a strong dependence of the radius estimators ($R_{\rm dep},
R_2$, etc., see Table~\ref{tab-cor}) on the system size Expr.~(\ref{correction}),
where the leading subdominant exponent
is estimated as $\nu\approx 0.33$.

It is well known~\cite{ms-PZ,Lee,nms-SBBS} that the fractal dimension
$D(x)$ can be found using the probability $P(r,N)$ for the N-th particle to
be deposited within a shell of width ${d}r$ at a distance $r$ from the
seed. The simplest and  most obvious form of the probability is
$P(r,N)=\frac{1}{\xi(N)}f(\frac{r-R_{\rm dep}(N)}{\xi(N)})$, where $f(y)$
is Gaussian. Practically, the Gaussian distribution can be obtained by
averaging over a large number of clusters. For the single-cluster
realization this function has some particular form reflecting the
details of the cluster growth as shown in Figure~\ref{probab}, in
which each local maximum is associated with an actively growing 
branch.

In~\cite{nms-SBBS} Somfai et al. computed $D(x)$ from the
Gaussian probability $P(r,N)$, and $D(x)$ with corrections to scaling
coincides well with the numerical results of Amitrano et
al.~\cite{ms-ACMZ}.  Somfai et all state that $D(x)$ tends to
a constant value as $N\to\infty$. In other words, Somfai et al.
advocate that multiscaling is transient and is an artifact of the finite
size of the DLA clusters.

In contrast, our numerical results demonstrate that (1)~there is no
evidence for the finite-size corrections with the exponent $\nu=0.33$,
(2)~$D(x)$ seems not tend to a constant, and (3)~it is not correct to 
use the Gaussian probability $P(r,N)$ to compute $D(x)$ for the DLA model.

Gaussian distribution does not reflect details of the DLA cluster
because it is the outcome of the 
the averaging over a large number of clusters. After such averaging
all details of the random nature of the growth of a particular cluster are
washed out, and the Gaussian distribution is just the result of the
central limit theorem. There are some indications by
Hastings~\cite{Hastings}, who computed DLA fractal dimension from field
theory, that the Gaussian distribution by itself is insufficient for
describing DLA
clusters, and some noise must be added to obtain a model
corresponding to the DLA model. To some extent, the local maxima in
Figure~\ref{probab} are due to that noise, in contrast to the
distribution averaged over the cluster ensemble, which is smooth.

Accordingly, $D(x)$ calculated from the averaged distribution would be
constant in the limit of large $N$. We can therefore say that multiscaling
comes from the fluctuations (natural noise) of the DLA cluster growth
process.

The fractal dimension $D(x)$ for different cluster sizes are shown in 
Figure~\ref{d_x} together with the results from~\cite{ms-ACMZ}. There is a 
notable maximum in $D(x)$ at around $x=1.4$, and its size does not change 
significantly with the number of particles in a cluster. We note that the 
position of the maximum can be found from the ratio of $(R_{\rm dep}-\xi)$ 
to $R_{\rm gyr}$. In our simulations, it equals $1.19$, which coincides 
well with the data in Figure~\ref{d_x}.

The maximum seems to occur because growth is mostly completed in the 
region $r<R_{\rm dep}-\xi$ , and the fractal dimension $D(x)$ decrease to 
the left of $x=(R_{\rm dep}-\xi)/R_{\rm gyr}$ in the actively growing 
region $D(x)$ because of the lack of particles there.  For small $x$, 
i.e., for $r<<R_{\rm gyr}$, no new particles are being added, and $D(x)$ 
is the same for different cluster sizes.

We note that the prediction of Somfai et al. for $D(x)$ with a system size 
$N=10^7$ is quite smaller than the one computed by us and presented in 
Figure~\ref{d_x}. We also estimate limit of the three curves plotted in 
Fig.~\ref{d_x}, taking limit of $N\rightarrow \infty$ at the fixed value 
of $x$, and plot result with the solid circles in Fig.~\ref{d_x}. Thus, 
our data does not demonstrate tendency of $D(x)$ to a constant value, and 
rather support our observation that the dynamic growth of the cluster is 
dominated by the active zone, and maximum in $D(x)$ reflects this nature 
of DLA. At the same time we have to note that accuracy of the data 
presented here is not enough for the final decision, and future 
investigation with the higher accuracy have to be still done.

\section{Summary} \label{sec-sum}

We have implemented modifications to the DLA algorithm that help us to 
grow large DLA clusters and test some recent claims about its properties. 
In our experiments, aggregates do not exhibit corrections to scaling laws. 
Nevertheless, the results show multiscaling properties. This means that 
there should be another way for such a clusters to appear. We tried to 
analyze the processes responsible for multiscaling. We will address this 
question in future research.

\acknowledgments

We appreciate Lev Barash for the careful reading of the manuscript, V.V. 
Lebedev for the valuable comments, and S.E. Korshunov for the constructive 
critics. Partial support of Russian Foundation for Basic Research is 
acknowledged.

\appendix \section{Probability to be alive on the birth circle} 
\label{sec-prob}

 The main result of this Appendix is expression~(\ref{prob-norm}). The 
same expression was obtained earlier~\footnote{We thank R. Ziff and 
anonymous referee for pointing us this papers. Authors of the present 
paper were not aware on that result prior the paper submission.} in 
Refs.~\cite{SSZ,KVMW}. We found our result is still worth publishing, since we 
derive it from a different point of view.

\begin{figure} \centering
\includegraphics[angle=0,width=\columnwidth]{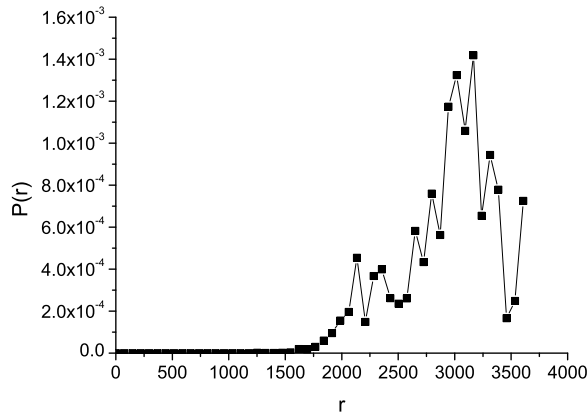}
\caption{Probability $P(r)$ for a particle to stick at the distance $r$, 
computed for a single DLA cluster.}
\label{probab}
\end{figure}

We consider a particle at some position $(r;\phi=0)$ outside
circle of radius $R_{\rm b}$, $r>R_{\rm b}$.
The particle moves randomly, and its size 
and
step is much smaller than $r$ and $R_{\rm b}$. The question is what is the
probability for a particle to intersect the circle $R_{\rm b}$ at the
angle
$\varphi'$. Clearly, for the particles walking from infinity,
i.e., $r>>R_{\rm b}$, that distribution is uniform in $[-\pi;\pi]$:

\begin{equation}
P(\varphi)d\varphi=\frac{1}{2\pi} d\varphi.
\label{prob-uni}
\end{equation}

\noindent The conformal map
\begin{equation}
w=R_{\rm b}\frac{zr+R_{\rm b}}{r+zR_{\rm b}}
\label{map}
\end{equation}

\noindent maps infinity to $(r,0)$ and the unit circle to the circle of
radius $R_{\rm b}$. Transformation~(\ref{map}) changes
probability~(\ref{prob-uni}) to the modulus of derivative of the
conformal map,

\begin{equation}
\frac{dz}{dw} = \frac{R_{\rm b}^2-r^2}{R_{\rm b}(r-w)^2}.
\label{map-der}
\end{equation}

\begin{figure} \centering
\includegraphics[angle=0,width=\columnwidth]{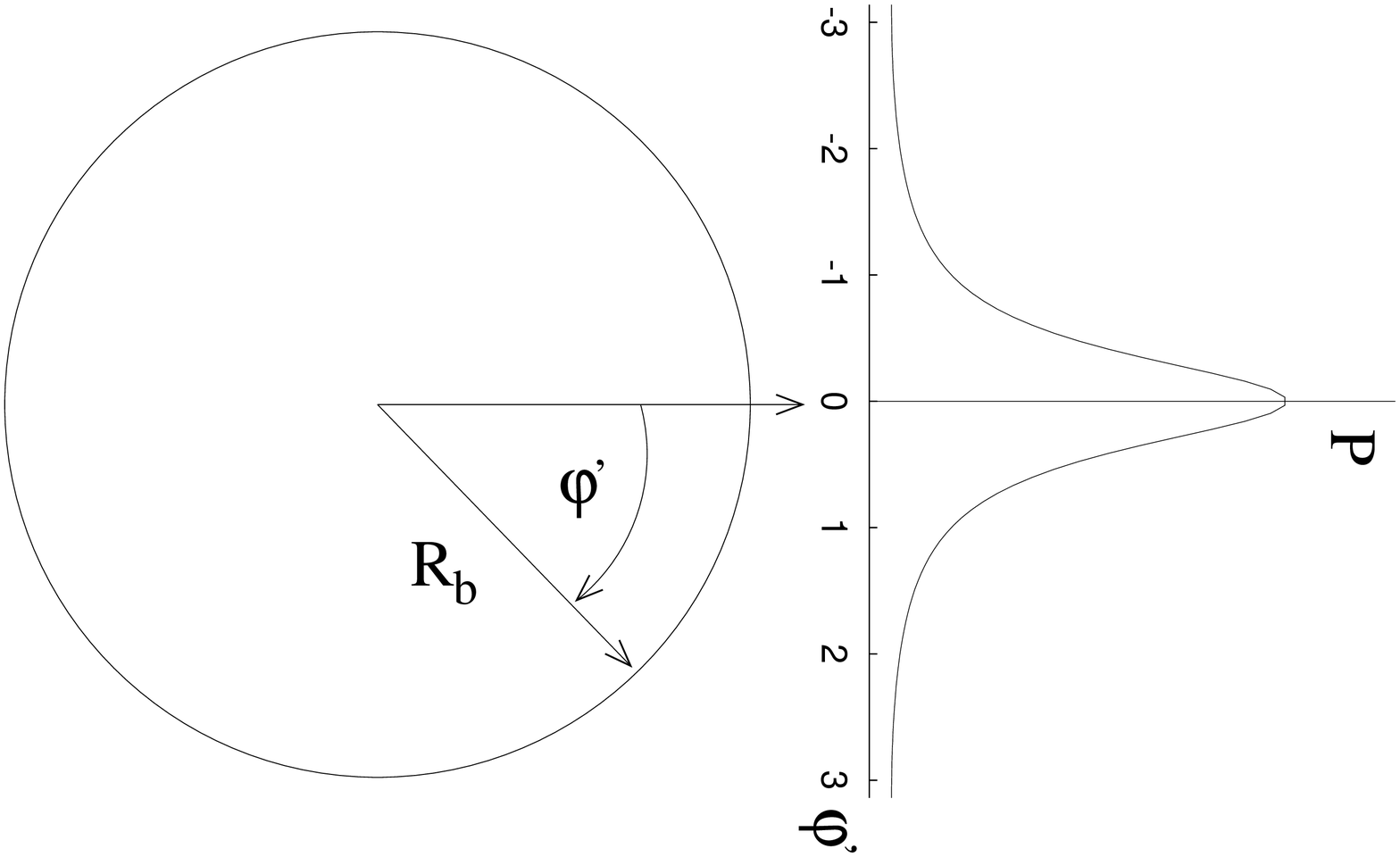}
\caption{Sketch of probability $P(\varphi')$ of 
the walk intersection of the circle
$R_{\rm b}$ with the initial position of the particle at $(r,0)$.}
\label{fig-prob}
\end{figure}

\noindent Substituting $w=R_{\rm b}\exp(i\varphi')$ in (\ref{map-der}), we
obtain the resulting probability
\begin{equation}
P(\varphi')=\frac{\rm const}{x^2-2x\;\cos\varphi'+1}
\label{prob-non}
\end{equation}

\noindent as a function of the ratio $x=r/R_{\rm b}>1$. This is a
probability for the particle beginning its walk at the point $(r,0)$
outside $R_{\rm b}$ to intersect circle $R_b$ at the angle $\varphi'$ 
(see  Figure~\ref{fig-prob}).

The constant in~(\ref{prob-non}) is associated with the probability for 
the particle to move to infinity. It can be identified using the analogy 
between the DLA model and Laplacian growth as was pointed 
in~\cite{Pietronero} for the dielectric breakdown model, which is a 
generalization of the DLA model.

The probability that the particle sticks somewhere to the cluster 
is proportional to the
electrical field at that point. We consider three circles of radii
$R_1$, $R_2$, and $R_3$, $R_1<R_2<R_3$. The external circles $R_1$ and
$R_3$ are set under the potential $\Phi=0$, and particles stick one of
them. The circle $R_2$ (place of a birth for particles)
is set with $\Phi=1$.  The solution of the Laplace equation
$\nabla^2\Phi=0$ with the above boundary conditions gives the distribution
of the potential and thus of the electrical field $\bf{E}$.
The probability for
the particle to stick on the circle $R_1$ would be
$P_1=\rm{const}*\int_{|r|=R_1}|\bf{E}|\,{d}^2r$ and 
similarly for $P_3$. The ratio of
these probabilities is:
\begin{equation}
\frac{P_3}{P_1}=\frac{\log \frac{R_2}{R_1}}{\log \frac{R_3}{R_2}}
\label{prob-ratio}
\end{equation}
\noindent The probability $P_3$ vanishes as $R_3 \to \infty$: $P_3\to 0$. 
This means that all particles starting on the birth radius should 
irreversibly collide with a cluster and the constant in~(\ref{prob-non})
 is easily found by normalizing the probability, 
$\rm{const}=(x^2-1)/2\pi$. The final expression for the probability 
\begin{equation} 
P(\varphi')=\frac{1}{2\pi}\frac{x^2-1}{x^2-2x\;\cos\varphi'+1} 
\label{prob-norm} 
\end{equation} 
\noindent provides correct values for both the limits $x\to \infty$ and 
$x\to 1$. The same expression for probability was described in~\cite{SSZ}, 
but it was obtained other way. Transformation
$$f(u)=2\arctan\left(\frac{x-1}{x+1}\tan(u\pi/2)\right)$$ 
maps uniformly distributed in interval $[-1,1]$ random number $u$ to 
random number with distribution~(\ref{prob-norm}).

\section{Memory organization}
\label{sec-mem}

Step 3 of the algorithm described in Section~\ref{sec-alg} essentially 
contains two routines. We must choose, first, the direction of the random 
walk and, second, the walk length. The direction of the walk is chosen 
uniformly in $[0;2\pi]$. Because the motion is uniform in direction, we 
can increase the step, but only if particle is far away from the cluster 
(in our simulations we choose that distance such that particle is more 
than five units away from the cluster, otherwise, its step is one unit of 
length). The length for the big step is chosen with the condition that the 
walk should not intersect any particles of the cluster. Therefore, the 
distance to the cluster $d_{\rm pc}$ is evaluated, and the step length is 
taken as that distance.

The reason for implementing variable step length is as follows. Most of 
the time the particle is moving far away from the cluster, and choosing a 
step length of the order of the particle distance from the cluster 
accelerates simulations.

To realize all the proposed improvements efficiently we must organize 
memory in a special manner. When the particle moves, we must know whether 
it collides with a cluster, and to check this, we must iterate over all 
particles. Such an approach is rather unreasonable and we must therefore 
restrict the number of particles to test. This is easily done by dividing 
the space into square cells, each about twenty units of length. Each cell 
saves information about particles that stuck in the region covered by it. 
We therefore need only check cells that are in the region of one step.

\begin{figure} \centering
\includegraphics[angle=0,width=\columnwidth]{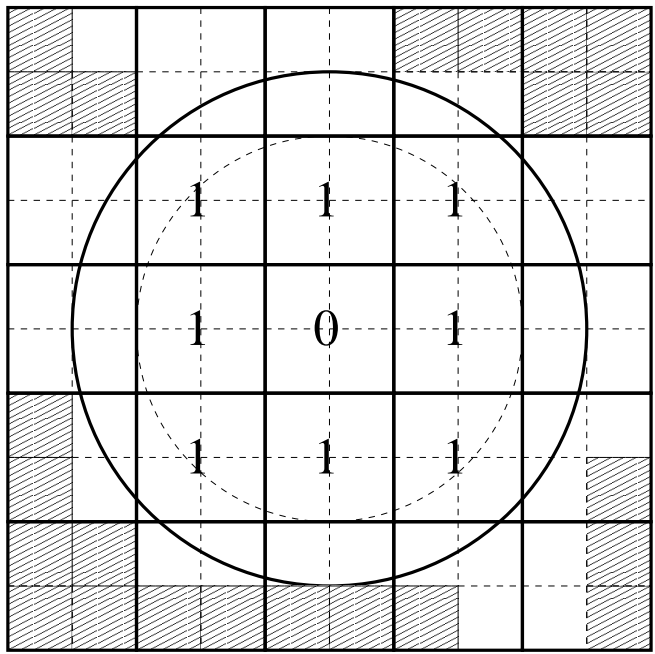}
\caption{Determination of the step length.}
\label{field}
\end{figure}

This model also improves the process of seeking the size of free space for 
the big step. To find the distance from current position to the cluster 
precisely is a difficult task, but it suffices in most cases to know it 
with an accuracy
 estimated from below, the size of one cell. Figure~\ref{field} shows how 
this is done. Cells are plotted in the picture as squares with bold lines. 
They are marked with numbers showing their distance from the particle 
location. For simplicity, cells with the same number are thought to be at 
the same distance from the particle. Occupied cells are shaded. In this 
example the particle is allowed to jump with the step length $L-2R$, where 
$L$ is the cell size, and $R$ the particle radius. The distance $L$ is the 
radius of the inscribed circle with a center somewhere in cell $0$ (in the 
worst case, the center lies on the border of cell $0$). This length should 
be reduced because cells save only particle centers and there could be a 
projection of a particle into another cell with a size of $R$. To simplify 
the algorithm we seek the step length only using the free/occupied cells 
picture. The first step is to check whether there are other particles in 
the cell we are now in, then we should check cells marked with 1, then 
marked with 2 (not shown in picture), and so on until we find an occupied 
cell.

During DLA growth, the intervals between branches increase notably, and 
the time to traverse all cells while seeking an occupied one also 
increases. To reduce the influence of such a process, Ball and 
Brady~\cite{BB-alg} developed a hierarchical memory model, where one cell 
is divided into smaller subcells and so on. This approach seems rather 
memory consumptive: in growing a large cluster, it would become a 
bottleneck of an algorithm.

The desired effect can be achieved in another way. Because the cell size 
 is much more bigger than the particle size, the distribution of free 
cells changes slowly, and process of seeking the maximum free space could 
be started not from the particle position but from the free line achieved 
in the previous search from this origin. To implement this we must save 
the value of free space around each cell. If this information is unknown, 
i.e., it is the first time to seek the maximum step length from the 
current cell, we should traverse all cells from the beginning; otherwise 
we start from the line previously saved and move to the center.

The cell size is chosen as follows. It should not be very small:  a small 
size results in high memory consumption and a rapid change of the 
distribution of free cells. On the other hand, its size restricts the 
precision of the length we find for the big step, i.e., there is a region 
of cell size near the cluster where particles can move only with a small 
steps. To avoid this last constraint, we use second-layer information. 
Like Ball and Brady, we divide the cells into smaller ones. To minimize 
memory consumption, we realize them as an 32 bit integer, where each bit 
shows whether the corresponding subcell is occupied. Each big cell can 
therefore be divided into not more than 25 subcells.

As mentioned above, the second-layer information is only used when the 
particle moves near the cluster, where the accuracy given by the first 
layer is insufficient. Figure~\ref{field} shows large cells (bold borders) 
and small cells. Using the first layer we can find that only the space 
inside the dashed circle is free. The second layer gives a more precise 
result: the particle can jump up to the bold circle.

\end{document}